\documentstyle [prl,preprint,aps] {revtex}

\begin{document}
\draft 
\preprint{Applied Physics Report No. 93-21}

\title{
  Non-Equilibrium Magnetization in a Ballistic Quantum Dot
}

\author{T. Swahn,$^{(1)}$ E. N. Bogachek,$^{(1,2)}$
Yu. M. Galperin,$^{(3)}$ M. Jonson,$^{(1)}$ and R. I. Shekhter$^{(1)}$}

\address{
     $^{(1)}$Department of Applied Physics,
     Chalmers University of Technology and G{\"o}teborg University,
     S-412 96 G\"oteborg, Sweden    \\
$^{(2)}$School of Physics, Georgia Institute of Technology,
  Atlanta, Georgia 30332\\
$^{(2)}$Department of Physics, University of Oslo, P.O. Box 1048 Blindern
0316 Oslo, Norway, \\ and A. F. Ioffe Institute, 194021 St. Petersburg, Russia
}

\maketitle

\begin{abstract}
We show that Aharonov-Bohm (AB) oscillations in the magnetic moment of
an integrable
 ballistic quantum dot can be destroyed by a time dependent magnetic
flux. The effect is due to a nonequilibrium population of perfectly coherent
electronic states. For real ballistic systems the equilibrization process,
which involves a special type of inelastic electron backscattering, can
be so ineffective, that AB oscillations are suppressed when the flux varies
with frequency $\omega\sim$ 10$^7$-10$^8$ s$^{-1}$. The
effect can be used to measure relaxation times for inelastic backscattering.
\end{abstract}

\narrowtext\noindent
The magnetic moment (and the associated persistent current) induced by a
magnetic flux in small
conductors is a dramatic manifestation of mesoscopic behavior.
While originally
predicted to appear in clean one-dimensional (1D) metallic rings
\cite{imry:91},
much of the recent discussion about persistent currents has focused on metallic
rings containing impurities \cite{levy:90,chan:91}. An important effect of the
inevitable impurity scattering in such systems is that `forbidden gaps' appear
in the spectrum of  quantized  electron energy levels when plotted
as a function of  magnetic flux. These gaps lead to an oscillatory
dependence of the single electron energies on flux and consequently to a
periodic magnetization  of  small  rings  and  dots
\cite{gang1}

The response to a magnetic flux that does not vary slowly in time may differ
from the quasi-static magnetization because of Landau-Zener macroscopic
tunneling through the disorder-induced  gaps \cite{gang2}.
In sufficiently `clean' systems such tunneling permits the energy
levels to take their free-electron values. For integrable ballistic 2D
structures formed
in gated semiconductor heterostructures, the
required minimum rate of change of flux, $\omega_{min}$, can
be low enough to be experimentally accessible \cite{rate}. Recent experiments
convincingly show Aharonov-Bohm (AB) oscillations in a 2D ballistic ring
\cite{mail:93} and the magnetization of
ballistic squares has also recently been measured \cite{Levy93}.
Even without any gaps in the energy spectrum a varying magnetic field may
result in a magnetization that oscillates with flux as observed. This is due
to a completely different mechanism involving a redistribution of electrons
between levels which are shifted up or down in energy by the changing flux.
The redistribution depends on inelastic relaxation processes that cannot
be associated with conventional one-phonon scattering. This is because the
relaxation of the induced magnetic moment requires simultaneously a large
momentum- and a small energy transfer. In this
Letter we point out that, as a result, the effective relaxation time $\tau$
is large and  one  can  expect  strong
non-equilibrium  effects in the magnetization.
The degree of non-equilibrium
depends on the rate of change of flux, $\omega$. If  $\omega_{\min} \ll \omega
\ll 1/\tau$  we find that the response of a pure quantum dot is quasi-static;
i.e.  
the magnetization takes on values corresponding to equilibrium at the
instantaneous magnitude of the flux. The variation --- both in magnitude and
sign --- of the equilibrium magnetic moment with magnetic flux $\Phi$ is
characterized by two distinctive  scales.  In addition to a flux variation on
the Aharonov-Bohm scale of $\Phi_0=hc/e$, the flux quantum,
there are oscillations on a smaller scale $\Phi_0/k_F a$ that gives a fine
structure to the flux dependence of the induced magnetic moment
\cite{avishai,mure:93} ($k_F$ is the Fermi wave vector, $a$ is the dot radius).

A time dependent magnetic flux tends to drive the system out of equilibrium.
Two distinct types of non-equilibrium effects appear. Even not
very far from equilibrium,
$\omega\tau{ }_\sim^<1/(k_F a)$,
the
fact that relaxation is not instantaneous has the effect of blurring the
fine-structure in the flux dependence of the magnetic moment
However, the oscillations of the Aharonov-Bohm type on the larger
scale of $\Phi_0$ remain  intact. Strong non-equilibrium behavior
appears when $\omega\tau$ becomes of order unity or  bigger;
then the  AB oscillations
are also  washed out and a non-equilibrium diamagnetic moment develops.
If magnetic flux increases (from zero) linearly with time, so does the moment
--- until time $\tau$ when inelastic backscattering is strong enough to
stabilize the diamagnetic response.

Below we will consider a `clean' quantum dot in a 2D electron gas, for which
the electron time-of-flight around the circumference, $a/v_F$, is the
shortest  time  of  the  problem,  $a/v_F   \ll   \omega^{-1},   \tau   \ll
\omega_{\min}^{-1}$.
Consequently, the electronic eigenstates remain well defined in a varying
external field
 and the system
stays coherent. The only result of a time dependent magnetic flux is a shift of
the electronic energy levels and hence a shifted
electron distribution in momentum space that
stays as sharp as the original one;
no broadening whatsoever appears due
to the time variation of the magnetic field.
The  non-equilibrium behavior is determined by the kinetics of the
energy level population, which depends on the interplay between the driving
magnetic field (time-scale $\omega^{-1}$)  and relaxation processes ($\tau$).
The unusual effect of a {\em lack} of scattering is that quantum oscillations
in the magnetization are {\em destroyed}. Mainly because the levels remain
sharp, there is a possibility for the oscillations to be {\em restored} by
inelastic scattering.

As an illustration it is useful to first discuss electrons confined to a
1D ring.
The eigenenergies
$E_m=E_0(m-\alpha)^2$ [$E_0={\hbar^2 /
2m^*a^2}$ and $\alpha\equiv\Phi/\Phi_0$] of a ring are sensitive to flux
and can be labelled by a magnetic quantum number $m=0, \pm 1, \pm2, \ldots$
In zero flux the population of $\pm m$-states is symmetric, as in Fig.~1a, and
there is no net magnetic moment. Increasing the flux $\alpha$ shifts the
states with negative (positive) quantum numbers to higher (lower)
energies as shown in Fig.~1b, and a net
magnetic moment appears. At the same time, a relaxation of the electron system
becomes possible through transitions between high energy $-m$ and low energy
$+m$-states (Fig.~1b). The interplay between the flux-driven shift of energy
levels and the `backflow' due to relaxation determines the kinetics of the
system.

Fig.~1 explicitly illustrates the main features of the relaxation
processes that
can  lead to an equilibration of the momentum distribution.
First, they have to be {\em inelastic} to compensate for the energy mismatch
that is a result of the quantization of energy levels. Secondly, it is
necessary
that they can provide the large momentum required to reverse the direction of
the azimuthal component of the electron momentum ($-m \rightarrow +m$).
We  emphasize that this type of `inelastic backscattering' does not play a role
in ordinary transport problems; it is a special feature of our system. We will
discuss possible mechanisms for inelastic backscattering later.

The above qualitative discussion is relevant also for a quantum dot as only
the azimuthal motion of the electron important for
generation and relaxation of a magnetic moment.
In the following we will discuss the quantum dot
in a  weak magnetic field, where
the quantized free electron energies  can be expressed as
\cite{dingle}
\begin{equation}
E_{m,n}=E_0\left[\gamma_{m,n} +2m\alpha
+\frac{\alpha^2}{3}\left(1+\frac{2(m^2-1)}{\gamma_{m,n}^2}\right)\right].
\label{spectrum}
\end{equation}
Here  $\gamma_{m,n}$ is the $n$:th root of the $m$:th Bessel
function, {\em i.e.} $J_m(\gamma_{m,n})=0$,
$n=1, 2,\ldots$ and $m=0, \pm 1, \pm 2,\ldots$
We can identify two different energy
scales in this spectrum. The smaller scale, $E_0$,
corresponds to the average spacing
between levels in zero flux; the larger scale,
$\Delta E=E_0 m \sim E_0 (k_F a)$
is set by the characteristic
shift of the energy levels when the flux is increased by $\Phi_0$.
The temperature
dependence of the fluctuations in magnetization is
related to these energy scales;
the fine structure oscillations start to disappear at temperatures of about
$T_1 = E_F/k_BN$ (here $N\sim (k_Fa)^2$  is the
number of particles in the dot and $E_F$ is the Fermi energy) while the large
scale oscillations only begin to vanish at the higher temperature
$T_2=E_F/k_B\sqrt{N}$.

In the quasi-static limit the induced
moment can be obtained  directly from the thermodynamic potential
of the system. One finds
\begin{equation}
M_{\text{eq}} = \frac{c}{S}\sum_{m,n} n_F(E_{m,n})
M_{m,n},
\label{ieq}
\end{equation}
where $M_{m,n}$ is the magnetic moment of a single quantum state ($m$,$n$),
\begin{equation}
M_{m,n} = -M_0
\left[ 2m + \frac{2\alpha}{3}\left(1+\frac{2(m^2-1)}{\gamma_{m,n}^2}\right)
\right].
\label{momentum}
\end{equation}
Here $M_0={\pi a^2 E_0 / \Phi_0}$,  $n_F(E)=\left[1+{\rm e}^{(E-\mu)/k_BT}
\right]^{-1}$
is the Fermi distribution function.
Results at different temperatures for the magnetic moment as a
function of
magnetic flux for the two cases of constant chemical
potential (dot connected to reservoir[s]) and constant number of particles
\cite{fixedN}
(isolated dot) are given in Fig.~2.
We note that the  flux dependence of the induced moment
appears qualitatively quite similar for the dot and a metallic ring. An
important quantitative difference is that the amplitude of the moment
fluctuations in the `clean'
dot is of order ${\pi a ev_F/c }$, rather than ${\pi ev_F\ell/c}$ as in a
`dirty' metallic ring when $\ell\ll a$. Induced moments
on this scale was recently observed in a ballistic {\em ring} system
\cite{mail:93}).

If the time variation of the magnetic flux is
not slow, essential differences appear; a nonequilibrium
distribution function, $f_{m,n}(t)$, has to replace the Fermi function
$n_F\left[E_{n,m}\left(\alpha(t)\right) \right]$ when the moment is
calculated from  Eq.~(\ref{ieq}).
To get explicit results we restrict ourselves to the simplest case for which
the relaxation time approximation applies. Here
\begin{equation}
\label{ke}
{\partial f_\gamma \over \partial t} = -
{f_\gamma(t) - n_F\left[E_\gamma\left( \alpha(t) \right) \right] \over
\tau_\gamma (t)},  \quad \gamma \equiv m,n,
\end{equation}
This approximation is valid at low temperatures when electrons decay
spontaneously  with a lifetime $\tau_\gamma(t)$. The latter depends on the
configuration of electronic levels $E_\gamma \left[\alpha (t) \right]$
and is therefore time dependent.

The induced moment for the
special case that the magnetic flux increases linearly with time,
$\alpha(t)= \omega t$, is shown in Fig.~3. For simplicity, we have assumed the
relaxation time $\tau$ to be time independent. This is the case when the main
source of relaxation is due to an exchange of electrons between the quantum dot
and its surrounding. We believe that this approximation gives a qualitatively
correct picture also in general.
Two non-equilibrium effects can be seen; small deviations from equilibrium
result in a smearing of the fine structure in the flux-dependence of
the moment. This happens when $\omega\tau \sim {1 / k_Fa}$ (left inset of
Fig.~3). Large-scale oscillations of the induced moment, which
correspond to the `usual' Aharonov-Bohm effect with a period of the order of
the flux quantum, are not affected. Drastic changes appear when the relaxation
time is so large that  $\omega\tau \sim 1$ or smaller. In this case the AB
oscillations disappear in favor of a diamagnetic current -- linearly dependent
on flux (time) -- until it saturates at some large flux (time). The value of
the  saturation moment can be readily obtained as
$M_{\text{sat}} \sim N M_0 \omega\tau$ .
 This result tells us that the saturation moment is limited only by a finite
relaxation time $\tau$ and can be very large in systems where relaxation  is
weak. Another interesting feature is the transient processes
appearing in the case of step-like changes in the flux-value. An example
of such a behavior is shown  in the right inset of Fig.~3. Here the
flux is (instantaneously) changed from one value to another, for which
the equilibrium induced moment is not very different. Nevertheless,
the initial diamagnetic  response is quite large and decays to
the new equilibrium value on the time scale $\tau$.

A number of mechanisms can be expected to relax a nonequilibrium momentum
distribution in the quantum dot.
If connected to a reservoir of particles
(fixed chemical potential), the reservoir acts as a sink for
energetic dot-electrons and a source of thermal electrons.
This
particle-exchange mechanism can be characterized by a relaxation time $\tau_e$.

Completely different relaxation mechanisms, associated with inelastic
backscattering, come into play if the quantum dot is isolated (fixed
number of particles). A possible mechanism for the required small energy- and
large momentum transfer
is simultaneous
scattering by impurities and phonons \cite{elel}.
Using a Green's function formalism \cite{AGD}, one finds an
order-of-magnitude estimate for the
scattering rate as $\tau_s^{-1}\approx {\hbar / E_F\tau_{\rm{imp}}
\tau_{\rm{ph}}}$. Here $\tau_{\rm{imp}}$ is the impurity- and
$\tau_{\rm{ph}}\approx {\Delta^3 / \hbar^3\omega_D^2}$ \cite{AGD} the
phonon relaxation time,
 $\omega_D$  is  the  Debye  frequency,  while  $\Delta$ is the transferred
 energy.
This energy is proportional to the magnetic flux;  for a transition between the
states $n,m_1 \rightarrow n,m_2$
one has $\Delta \approx 2E_0 \alpha |m_1-m_2|$.
Using these estimates and typical parameters for semiconductor
nanostructures one gets $ \tau_s^{-1}\approx (10^7-10^8) \alpha^3$.  We
stress that {\em any} scattering mechanism leads to a
time-dependent relaxation rate because of the time dependence of $\Delta$,
the flux-dependent energy difference between initial and final state.
Hence, contrary to the
particle-exchange mechanism, we expect inelastic backscattering to cause a
non-exponential relaxation of the non-equilibrium magnetic moment. The total
relaxation rate can be estimated as
$\tau^{-1}=\tau_e^{-1} + \tau_s^{-1}$.
We believe that the relative importance of the two mechanisms can be controlled
in structures where the coupling between the quantum dot and adjacent
reservoirs can be varied by means of a gate voltage.
Among other possible mechanisms for inelastic backscattering, one can mention
inelastic   scattering caused by  atomic  two-level  systems   (see \cite{gkk}
 for a  review) and  by electronic
two-level systems created by close pairs of filled and empty donors in a doped
region of the structure (always present in semiconductor heterostructures).

In   conclusion  we  have  shown  that  in  a  varying  magnetic  flux  the
magnetization of
 a 2D ballistic quantum dot is very sensitive to the conditions of
relaxation in the system.
In contrast to the usual destructive role played
by inelastic scattering in mesoscopic phenomena, here inelastic scattering
{\em restores} an Aharonov-Bohm type of quantum oscillations in the
magnetization. In the absence of such relaxation, strong non-equilibrium
behavior suppresses these oscillations in favor of large diamagnetic
moments which are
determined by flux- rather than Landau level quantization as in bulk
materials. A special type of inelastic backscattering is responsible for
relaxation in the case of an isolated dot, and determines the maximum
(saturation) value of the non-equilibrium diamagnetic moment in the case of
a magnetic flux which increases linearly with time. By monitoring the transient
behavior of the induced moment as the magnetic field is switched from one
value to the other we propose it might be possible to measure the
characteristic time of inelastic  backscattering estimated to be
of the order of 10$^{-8}$ - 10$^{-7}$ seconds.

This work was supported by the Swedish Royal Academy of Sciences,
the Natural Science Research Council, and by NUTEK. We acknowledge
enlightening discussions with L. Glazman.  E.B.
acknowledges the hospitality of the Department of Applied Physics, CTH/GU.


\begin{figure}
\caption{
Schematic dependence of the energy eigenvalue $E_{m,n}$ on the magnetic
quantum number $m$ for fixed $n$ (corresponding to a ring-shaped part
 of the dot):
(a) occupied (filled circles) and unoccupied (empty circles)
quantum states at zero magnetic field, (b) energies of
the same states at finite field. To reach equilibrium
the system has to relax by inelastic backscattering events involving small
energy- and large momentum transfer (indicated by arrow; see text).
}
\end{figure}

\begin{figure}
\caption{
(a) Magnetic moment vs. normalized magnetic flux $\alpha=\Phi/\Phi_0$
in a ballistic quantum dot with fixed chemical potential $\mu$
calculated at various temperatures. Units of moment, $M_0=\pi a^2 E_0/\Phi_0$,
and temperature, $E_0/k_B$, contain the quantum unit of flux $\Phi_0$ and the
average spacing between energy levels $E_0$.
The fine structure in the flux dependence of the moment disappears at
$T\sim 1$,
whereas the Aharonov-Bohm oscillations of period unity ($\Phi_0$) persist
until $T\sim k_Fa$ ($a$ is the dot radius).
(b) Zero-temperature calculations for fixed
$\mu$ and for fixed number of particles give qualitatively similar results.
}
\end{figure}

\begin{figure}
\caption{
Magnetic moment in a ballistic quantum dot vs. a normalized magnetic flux
that grows linearly with time, $\alpha(t)=\omega t$, from $t=0$.
The parameter $\omega\tau$, where $\tau$ is the relaxation
time, measures the rate of change of flux. The top left panel
shows that the fine structure in the flux dependence of the moment is
smeared when $\omega\tau\sim 1/k_Fa$ ($a$ is the dot radius).
For larger values
of $\omega\tau$ a large diamagnetic moment is proportional to
$\alpha(t)\propto t$
until it saturates at $\alpha\sim \omega\tau$. The top right panel shows the
current response to a sudden change of flux (cf. top left panel).
By monitoring how
the current relaxes towards a new equilibrium value, $\tau$ could be measured.
}
\end{figure}


\begin{references}
\bibitem{rate}
In modulation doped AlGaAs heterostructures
the short wavelength potential fluctuations responsible for the
energy gaps are exponentially suppressed. Structures with
wide spacer regions and smooth geometry can hence be expected to be ballistic
(more precisely integrable). Detailed estimates of $\omega_{min}$ will be
published elsewhere.
\bibitem{fixedN}
At zero temperature, when $E_F$ unambigously separates filled and
unfilled energy levels, it is straightforward to let $E_F$
fluctuate to keep the  number of particles fixed.
\bibitem{elel}
Note that electron-electron scattering is not relevant because it cannot
change  the total magnetic moment of the system and it will not result
in moment  relaxation.
\end{references}
\end{document}